\documentclass[12pt]{article}
\usepackage{amssymb,amsfonts}
\usepackage{graphics,psboxit,amsmath}
\usepackage{subfigure}
\usepackage{graphicx}
\usepackage{verbatim}

\def\hybrid{\topmargin -30pt    \oddsidemargin 0pt 
        \headheight 0pt \headsep 0pt
        \textwidth 6.25in       
        \textheight 9.5in       
        \marginparwidth .875in
        \parskip 5pt plus 1pt   \jot = 1.5ex}

\hybrid

\def\baselinestretch{1.2}

\catcode`\@=11

\def\marginnote#1{}
\newcount\hour
\newcount\minute
\newtoks\amorpm
\hour=\time\divide\hour by60 \minute=\time{\multiply\hour by60
\global\advance\minute by-\hour}
\edef\standardtime{{\ifnum\hour<12 \global\amorpm={am}%
        \else\global\amorpm={pm}\advance\hour by-12 \fi
        \ifnum\hour=0 \hour=12 \fi
        \number\hour:\ifnum\minute<10 0\fi\number\minute\the\amorpm}}
\edef\militarytime{\number\hour:\ifnum\minute<10 0\fi\number\minute}

\def\draftlabel#1{{\@bsphack\if@filesw {\let\thepage\relax
   \xdef\@gtempa{\write\@auxout{\string
      \newlabel{#1}{{\@currentlabel}{\thepage}}}}}\@gtempa
   \if@nobreak \ifvmode\nobreak\fi\fi\fi\@esphack}
        \gdef\@eqnlabel{#1}}
\def\@eqnlabel{}
\def\@vacuum{}
\def\draftmarginnote#1{\marginpar{\raggedright\scriptsize\tt#1}}

\def\draft{\oddsidemargin -.5truein
        \def\@oddfoot{\sl preliminary draft \hfil
        \rm\thepage\hfil\sl\today\quad\militarytime}
        \let\@evenfoot\@oddfoot \overfullrule 3pt
        \let\label=\draftlabel
        \let\marginnote=\draftmarginnote
   \def\@eqnnum{(\theequation)\rlap{\kern\marginparsep\tt\@eqnlabel}%
\global\let\@eqnlabel\@vacuum}  }

\def\draft2{
        \def\@oddfoot{\sl preliminary draft \hfil
        \rm\thepage\hfil\sl\today\quad\militarytime}
        \let\@evenfoot\@oddfoot \overfullrule 3pt
        \let\label=\draftlabel
        \let\marginnote=\draftmarginnote
   \def\@eqnnum{(\theequation)\rlap{\kern\marginparsep\tt\@eqnlabel}%
\global\let\@eqnlabel\@vacuum}  }

\def\preprint{\twocolumn\sloppy\flushbottom\parindent 2em
        \leftmargini 2em\leftmarginv .5em\leftmarginvi .5em
        \oddsidemargin -.5in    \evensidemargin -.5in
        \columnsep .4in \footheight 0pt
        \textwidth 10.in        \topmargin  -.4in
        \headheight 12pt \topskip .4in
        \textheight 6.9in \footskip 0pt
        \def\@oddhead{\thepage\hfil\addtocounter{page}{1}\thepage}
        \let\@evenhead\@oddhead \def\@oddfoot{} \def\@evenfoot{} }

\def\numberbysection{\@addtoreset{equation}{section}
        \def\theequation{\thesection.\arabic{equation}}}

\def\underline#1{\relax\ifmmode\@@underline#1\else
        $\@@underline{\hbox{#1}}$\relax\fi}

\def\titlepage{\@restonecolfalse\if@twocolumn\@restonecoltrue\onecolumn
     \else \newpage \fi \thispagestyle{empty}\c@page\z@
        \def\thefootnote{\fnsymbol{footnote}} }

\def\endtitlepage{\if@restonecol\twocolumn \else \newpage \fi
        \def\thefootnote{\arabic{footnote}}
        \setcounter{footnote}{0}}  

\catcode`@=12 \relax

\def\figcap{\section*{Figure Captions\markboth
        {FIGURECAPTIONS}{FIGURECAPTIONS}}\list
        {Figure \arabic{enumi}:\hfill}{\settowidth\labelwidth{Figure
999:}
        \leftmargin\labelwidth
        \advance\leftmargin\labelsep\usecounter{enumi}}}
 \relax
\def\tablecap{\section*{Table Captions\markboth
        {TABLECAPTIONS}{TABLECAPTIONS}}\list
        {Table \arabic{enumi}:\hfill}{\settowidth\labelwidth{Table
999:}
        \leftmargin\labelwidth
        \advance\leftmargin\labelsep\usecounter{enumi}}}
 \relax
\def\reflist{\section*{References\markboth
        {REFLIST}{REFLIST}}\list
        {[\arabic{enumi}]\hfill}{\settowidth\labelwidth{[999]}
        \leftmargin\labelwidth
        \advance\leftmargin\labelsep\usecounter{enumi}}}
 \relax
\makeatletter
\newcounter{pubctr}
\def\publist{\@ifnextchar[{\@publist}{\@@publist}}
\def\@publist[#1]{\list
        {[\arabic{pubctr}]\hfill}{\settowidth\labelwidth{[999]}
        \leftmargin\labelwidth
        \advance\leftmargin\labelsep
        \@nmbrlisttrue\def\@listctr{pubctr}
        \setcounter{pubctr}{#1}\addtocounter{pubctr}{-1}}}
\def\@@publist{\list
        {[\arabic{pubctr}]\hfill}{\settowidth\labelwidth{[999]}
        \leftmargin\labelwidth
        \advance\leftmargin\labelsep
        \@nmbrlisttrue\def\@listctr{pubctr}}}
 \relax
\makeatother

\def\be{\begin{equation}}
\def\ee{\end{equation}}
\def\ba{\begin{eqnarray}}
\def\ea{\end{eqnarray}}

\def\del{\partial}

\def\a{\alpha}

\def\b{\beta}

\def\g{\gamma}

\def\d{\delta}

\def\th{\theta}
\def\Th{\Theta}
\def\m{\mu}
\def\n{\nu}
\def\om{\omega}
\def\Om{\Omega}
\def\l{\lambda}
\def\L{\Lambda}
\def\s{\sigma}

\def\cN{{\cal N}}

\def\no{\noindent}

\def\qq{\qquad}

\def\IR{\relax{\rm I\kern-.18em R}}

\def \ov {\over}

\def\const{{\rm const.}}

\begin{document}


\renewcommand{\theequation}{\thesection.\arabic{equation}}
\csname @addtoreset\endcsname{equation}{section}

\newcommand{\eqn}[1]{(\ref{#1})}
\begin{titlepage}
\begin{center}


\begin{flushright}
June 2008 \hfill
LPTENS-08/37 \\
CPHT-RR 043.0608
\end{flushright}

\vskip .5in

{\Large \bf Stability issues with baryons in AdS/CFT}

\vskip 0.5in

{ \bf Konstadinos Sfetsos}$^{1,2,3}$\phantom{x} and\phantom{x} {\bf Konstadinos
Siampos}$^{2,3}$ \vskip 0.1in

$^1$ {\it Laboratoire de Physique Th\'eorique de l'Ecole Normale Sup\'erieure}, CNRS–-UMR 8549\\
24 rue Lhomond, 75231 Paris Cedex 05, France

$^2$ {\it Centre de Physique Th\'eorique, Ecole Polytechnique}, CNRS – UMR 7644\\
91128 Palaiseau, France

$^3$ {\it Department of Engineering Sciences, University of Patras},\\
26110 Patras, Greece

\vskip .1in


{\footnotesize {\tt E-mail:\{sfetsos,ksiampos\}$@$upatras.gr}}\\

\end{center}

\vskip .4in

\centerline{\bf Synopsis}

\no
We consider baryon vertices within the gauge/gravity correspondence
for a class of curved backgrounds.
The holographic description based on the ${\cal N}=4$ SYM theory for $SU(N)$ allows
classical solutions representing
bound states of $k$-quarks with $k$ less than or equal to $N$.
We construct the corresponding classical configurations and perform a stability analysis.
We present the details for the theory
at the conformal point and at finite temperature and show that there is a critical value of $k$,
below which there is instability. This may also arise when the baryon reaches
a critical size.
We also extend our treatment to magnetically charged baryon vertices.

\vfill

\end{titlepage}
\vfill \eject


\tableofcontents

\def\baselinestretch{1.2}
\baselineskip 20 pt
\no

\section{Introduction}

The basic construction in use to understand baryons in the context of the gauge/gravity correspondence
\cite{adscft} involves $N$ heavy external quarks on the boundary of $AdS_5$
at the end points of  $N$ fundamental strings terminating at a vertex
on the interior of $AdS_5$.
Each string contributes a unit of charge, so that for the total charge to be conserved
at the vertex, it was proposed in \cite{Witten:1998} that a $D5$-brane wrapped on
the internal $S^5$-sphere should be placed there.
Due to its presence, a coupling via a Wess--Zumino term to the
Dirac--Born--Infeld (DBI) electric field strength ensures that the
total charge indeed is preserved.
This idea was actually applied in calculating the
classical solution corresponding to such a configuration in \cite{Brandhuber:1998,Imamura1}
and involved generalizations of
techniques developed for the heavy quark-antiquark system at the conformal point of
$\cN=4$ SYM for $SU(N)$ in \cite{maldaloop}.

\no
In recent works \cite{ASS1,ASS2,SS} we have extensively
studied within
the gauge/gravity correspondence, the stability of interaction potentials
between a heavy quark and anti-quark in mesons, as well as between
a quark and monopole, for a general class of backgrounds.
A general outcome of these works is that a stability analysis
is very important in establishing the physical parameter space of the theory which is typically
reduced in comparison with that allowed from simply the requirement that the
classical solutions exist.
A second lesson is that in cases of multibranch solutions the point where instability
begins does not always coincide with the point where a solution bifurcates.
It seems that this feature requires
the coexistence in the classical configuration of fundamental as well as of solitonic objects.
In particular, this
was the case of the potential for the quark-monopole pair
in which a string junction was used that requires precisely
a fundamental-, a $D$- and a dyon-string.

\no
The case of baryons we focus in this paper presents a new challenge and requires
special care. In particular, at a technical level it requires a multistring configuration with $N$
strings. In addition, the vertex where these meet in the bulk of $AdS_5$ is not a mathematical point,
as in the case of a string junction, but it is a place where a solitonic object, the $D5$-brane
wrapped on the internal $S^5$, is placed.
A further challenge, that initially prompted our analysis, is that there have been
configurations with $5N/8  < k < N$ quarks \cite{Brandhuber:1998,Imamura1}
that are perfect as classical solutions to the equations of motion, besides the
expected ones with $k=N$.
This fact is against physical intuition and experimental results
that, bound states of quarks and antiquarks are singlets of the gauge group. Of course in the case
at hand we deal with a maximally supersymmetric theory, but one would like to show that a better
bound for $k/N$ can be found, if not $k=N$ exactly.
It is the purpose of the present paper to address these and related issues.

\no
The organization of this paper is as follows: In section 2 we formulate baryons for
a general class of
backgrounds. In particular, we derive general formulas for the baryon binding energy for the
case of quarks uniformly distributed on a spherical shell when forming the baryon.
In
section 3 we perform a stability analysis under small fluctuations and
derive their equations of motion as well as the boundary and matching
conditions at the boundary of $AdS_5$ and at the baryon vertex.
In section 4 we present
examples of baryon configurations
within the gauge/gravity correspondence, using the conformal and finite temperature
 backgrounds.
We will show that stability restricts the value of $k$ to be larger than a critical value
even in the conformal case (a higher than $5N/8$ value).
Also even a part of the energetically favorable branch of
multi-branched potentials should be disregarded as unstable.
In section 5 we extend our treatment to the case of magnetically charged baryon vertices
and show that
stability requires un upper bound for the instanton number, as compared to the 't Hooft coupling,
associated with a self-dual gauge field.
Finally, in section 6, we summarize our results and discuss some future directions.

\section{The classical solution}
\label{sec-2}

In this section we develop and present the general setup of the
gauge/gravity calculation of the potential energy of a
baryon consisting of heavy static quarks.
Since the underlying dual theory is ${\cal N}=4$ SYM for $SU(N)$, these computations
involve $N$ strings as well a $D5$--brane wrapped on the internal $S^5$--sphere.
Certain elements of the computation will be based on the well known techniques
used for the calculation of the potential of a  heavy quark-antiquark pair
\cite{maldaloop} as generalized beyond the conformal point in \cite{bs},
as well as the concept of the baryon vertex \cite{Witten:1998}.
Our classical construction
will essentially generalize that for the conformal case in \cite{Brandhuber:1998,Imamura1} to
a more general class of backgrounds.

\no
We consider diagonal metrics of Lorentzian signature of the form
\ba
\label{2-1}
ds^2 = G_{tt} dt^2 + G_{xx}(dx^2 + dy^2 + dz^2)+G_{uu}du^2 + R^2d\Om_5^2 \ ,
\ea
where $x,y$ and $z$ denote cyclic coordinates and $u$ denotes the radial direction playing the
r\^ole of an energy scale in the dual gauge theory. It extends
from the UV at $u \to \infty$ down to the IR at some minimum value
$u_{\rm min}$ determined by the geometry.

\no
It is convenient to introduce the functions
\ba
\label{2-2}
f(u) = - G_{tt}G_{xx}\ ,\qq g(u)=- G_{tt} G_{uu}\ , \qq h(u)=G_{yy}G_{uu} \ .
\ea
In the conformal limit we can approximate the metric by that for
$AdS^5\times S^5$ with radii ($\a'=1$) $R=(4\pi g_s N)^{1/4}$,
with $g_s$ being the string coupling. In this limit the leading order expressions are
\ba
\label{2-3}
 f(u)\simeq u^4\ ,\qq g(u)\simeq 1\ ,\qq h(u)\simeq 1\ .
\ea
A baryon is by definition a bound state of $N$-quarks which form
the completely antisymmetric representation of $SU(N)$.
However, within the gauge/gravity correspondence,
we can in general construct a bound state of
$k$-quarks with $k<N$. This bound state consists of a $D5$-brane located in the
bulk and $k$-external quarks at the boundary of $AdS_5$.
Each quark is connected with a string to the $D5$-brane.
There are also $N-k$ straight strings
that go from the $D5$-brane straight up at $u_{\rm min}$.
This description is depicted in Fig. 1. We will take $N,k\gg 1 $, such that the ratio
$k/N$ is held finite.

\no
At this point we have to choose a profile for the quark distribution inside the baryon.
We expect a radially symmetric distribution and
for simplicity we take it to be a uniform distribution on a spherical shell
of radius $L$. We will briefly comment on other choices and on how
the preferable distribution could
be determined in the last section.
Also note the entire description is valid
only if the endpoints of the $N$-strings are uniformly distributed on the
internal
$S^5$-sphere so that the latter is not deformed and the probe
approximation still holds (for a discussion on this point see \cite{Imamura1}).
Such a choice completely breaks supersymmetry \cite{Imamura2,Callan:1998iq}.

\begin{figure}[!t]
\begin{center}
\begin{tabular}{cc}
\includegraphics[height=10cm]{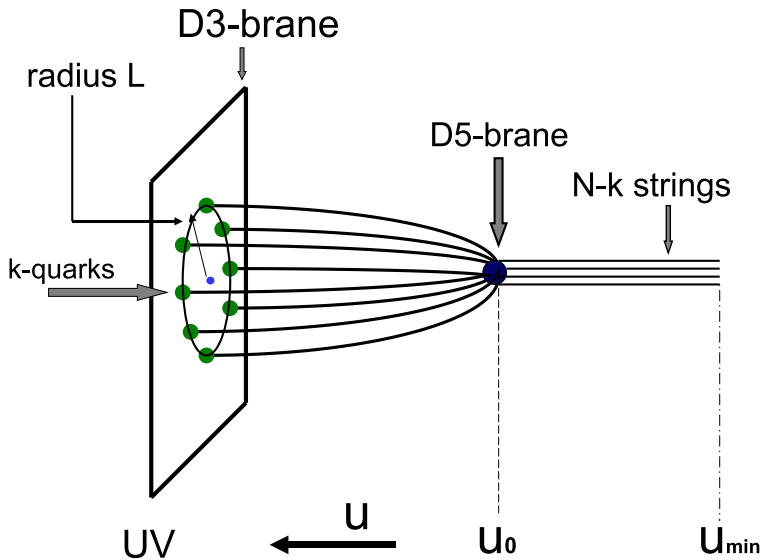}
\end{tabular}
\end{center}
\vskip -2.5 cm \caption{A baryon configuration with $k$-external quarks placed on a
spherical shell of radius $L$
at the boundary of AdS space, each connected to the $D5$-brane which is wrapped on an
$S^5$ located at $u=u_0$ and $N-k$ straight
strings ending at $u_{\rm min}$.}
\end{figure}

\no
Within the gauge/gravity correspondence,
the binding potential energy of the baryon is given by
\ba
\label{2-4}
e^{-{\rm i} E T} = e^{{\rm i} S_{\rm cl}}\ ,
\ea
where $S_{\rm cl}$ is the classical action of the above configuration.
\no
This action consists of three terms,
a Nambu--Goto action for each of the $N$--strings,
a Dirac--Born--Infeld action for the $D5$-brane and
a Wess--Zumino term between the $N$-strings and the $D5$-brane
\ba
\label{2-5}
&&S_{NG}[C] =  {1\ov 2 \pi} \int_C d \tau d \sigma \sqrt{-\det g_{\a \b}}\ ,
\nonumber\\
&& S_{D5}={1\ov (2\pi)^5 g_s}\int_M d^6x \sqrt{-h_{ij}-2 \pi F_{ij}}\ ,
\\
&&S_{WZ}={1\ov (2\pi)^5 g_s}\int_M d^6x\ C^{(4)}\wedge F\ ,\qq M={\mathbb{R}\times S^5}\ ,
\nonumber
\ea
where $g_{\a \b},h_{ij},F_{ij}$ are the induced metrics for the string, $D5$--brane and
Born--Infeld field strength, respectively. We will consider the simplest case
where the DBI--action is just given by its induced metric and
the presence of the Wess--Zumino term acts as a charge conservation term \cite{Witten:1998}.
However, there are cases of non-trivial $F_{ij}$ \cite{Lozano:2006} to which we will return later
in section 5.
We have not also included the contribution of the antisymmetric NS-NS two-form
due to the fact that it is not present in our examples.

\no
We first fix reparametrization invariance for each string by choosing
\ba
\label{2-6} t=\tau \ ,\qq u=\s \ .
\ea
For static solutions
we consider the embedding of the $S^2$--sphere on the $D3$--brane in spherical
coordinates $(r,\th,\phi)$
\ba
\label{2-7} r = r(u)\ ,\qq \phi, \ \th =\const\ ,\qq \hbox{$S^5$--angles} = \const\ ,
\ea
supplemented by the boundary condition
\ba
\label{2-8} u \left(L \right)=\infty\ .
\ea
Then, the Nambu--Goto action reads
\ba
\label{2-9} S = - {T\ov 2 \pi} \int du\sqrt{ g(u) + f(u) r^{\prime 2}}\ ,
\ea
where $T$ denotes time and the prime denotes a
derivative with respect to $u$. From the Euler--Lagrange equations of motion
we obtain
\ba
\label{2-10}{f r_{\rm cl}^\prime \ov \sqrt{g+ f r_{\rm cl}^{\prime 2}}}
=  f_1^{1/2}\qq \Longrightarrow
\qq r_{\rm cl}^\prime =  {\sqrt{f_1F}\ov f}\ ,
\ea
where $u_1$ is the value of $u$ at the turning point of each string,
$f_1 \equiv f(u_1),\ f_0\equiv f(u_0)$ and
\ba
\label{2-11} F = {g f \ov f - f_1}\ .
\ea
The $N-k$ straight strings which extend from the baryon vertex to $u=u_{\rm min}$
are straight, since $r=\const$ is always a solution of the equations of
motion (with $f_1=0$).
In the examples to follow $u_{min}\leqslant u_1\leqslant u_0$ and $f(u)$ is an increasing function.
Integrating \eqn{2-10}, we can express the radius of the spherical shell as
\ba
\label{2-12} L =\sqrt{f_1}\int_{u_0}^{\infty} d u {\sqrt{F}\ov f}\ .
\ea

\no
Next we fix the reparametrization invariance for the wrapped $D5$-brane by choosing
\ba
\label{2-13} t=\tau\ ,\qq \th_a=\s_\a\ , \quad \a=1,2,\dots,5\ .
\ea
For the ansatz given above, the DBI action reads
\ba
\label{2-14}
S_{D5}={TNR\ov 8\pi}\sqrt{-G_{tt}}\Big{|}_{u=u_0}\ ,
\ea
where we have also used that the volume of the unit-$S^5$ is $\pi^3$.

\no
Finally,
inserting the solution for $r_{\rm cl}^\prime$ into \eqn{2-9},
subtracting the divergent self-energy contribution of disconnected
worldsheets and adding the energy of the D5--brane \eqn{2-14},
we write the binding energy of the baryon as
\ba
\label{2-15}
E = {k \ov 2\pi}\left\{\int_{u_0}^\infty d u \sqrt{F}
- \int_{u_{\rm min}}^\infty d u \sqrt{g}
+{1-a\ov a} \int_{u_{\rm min}}^{u_0} d u \sqrt{g} +
{R\sqrt{-G_{tt}}\ov 4 a }\ \bigg|_{u=u_0} \right\}\ ,
\ea
where
\be
a\equiv {k\ov N}\ ,\qq 0< a \leqslant 1\ .
\ee

\no
The expressions for the length and the energy, \eqn{2-12} and \eqn{2-15} depend on the arbitrary
parameter $u_1$ which should be expressed in terms of the baryon vertex position $u_0$.
The most convenient way to find this
is to impose that the net force at the baryon vertex is zero.
Following an analogous  procedure to that for the string junction in \cite{SS}, we temporarily change
gauge to $r=\s$ and from the boundary terms arising in deriving the classical equations of motion
we find that for a uniform distribution of the quarks on a spherical shell, we get the relation
(reducing to that in \cite{Brandhuber:1998} for the conformal case)
\ba
\label{2-16}
&&\cos\Th = {1-a\ov a} + {R\ov 4a\sqrt{g}}\del_u\sqrt{-G_{tt}}\Big{|}_{u=u_0}\ ,
\nonumber\\
&& \cos\Th =  \sqrt{1-f_1/f_0}\ ,
\ea
where $\Th$ is the angle between each of the
$k$-strings and the $u$-axis at the baryon vertex
and which determines $u_1$ in terms of $u_0$.
The above relation can also be interpreted as the no-force condition for its
$u$-component \cite{Imamura1}.
The term on the left hand side of \eqn{2-16} is due to the strings attached to the boundary at the
UV. The force they exert on the baryon vertex is balanced by the forces exerted
on that by the straight strings and the tendency of the $D5$-brane to move
towards $u_{\rm min}$, represented
by the first term and the second term on the right hand side of \eqn{2-16} (the second
term is indeed positive in our examples).
Hence, if $a$ becomes small enough we expect that this condition might not be possible to
satisfy.
As we will see in the examples to follow, \eqn{2-16} has a solution
for a parametric region of $(a,u_0)$.
However, in order to isolate parametric regions
of physical interest a stability analysis of the classical solution should be performed, which,
as we will see,
further restricts the allowed region.

\section{Stability analysis}
\label{sec-3}
We now turn to the stability analysis of the above classical configuration,
aiming at isolating the physical parametric regions.
This involves string fluctuations which fall
into the general results analyzed in \cite{ASS1,ASS2} for a single string.
However,
the quantitative results are quiet different, due to the relative orientation of the strings, being
distributed on a sphere, and the different boundary conditions
that will be imposed here.

\subsection{Small fluctuations}

We parametrize the string fluctuations about the classical solution,
by perturbing the embedding according to
\ba
\label{3-1}
r = r_{\rm cl}(u) + \d r (t,u)\ ,\qq \th =\th_0 + \d\th (t,u)\ ,\qq \phi = \phi_0+\d\phi (t,u)\ .
\ea
\no
We have kept the gauge choice \eqn{2-6} unperturbed by using worldsheet
reparametrization invariance.
We then calculate the Nambu--Goto action for this ansatz
and we expand it in powers of the fluctuations.
The resulting expansion
for the quadratic fluctuations for each one of the $k$-strings
can be written as
\ba
\label{3-2}
\!\!\!\!\!\!\!\!\!\!\!\! S_2 &=& - {1 \ov 2\pi} \int
dt du \biggl[ {g f \ov 2 F^{3/2}}\ \d r^{\prime 2} - {h \ov 2 F^{1/2}}\
\d \dot{r}^2+{f r_{\rm cl}^2 \ov 2 F^{1/2}}
\ \d\th^{\prime 2} - {h r_{\rm cl}^2 F^{1/2} \ov 2 g}\  \d \dot{\th}^2 \nonumber\\
&&\qq\qq\qq + {f r_{\rm cl}^2\sin^2\th_0 \ov 2 F^{1/2}}\
\d\phi^{\prime 2} - {h r_{\rm cl}^2\sin^2\th_0  F^{1/2} \ov 2 g}\ \d \dot{\phi}^2\biggr]\ .
\ea
Using the Euler--Lagrange equations and the ansatz
\ba
\label{3-3}
\d x^\m (t,u) = \d x^\m (u) e^{-\imath \om t}\ ,\qq x^\m=r,\th,\phi\ ,
\ea
we find for each of the $k$--strings the linearized equations for
the longitudinal and transverse fluctuations
\ba
\label{3-4}
&&\left[ {d \ov du} \left( {g f \ov F^{3/2}} {d \ov du} \right)
+ \omega^2 {h \ov F^{1/2}} \right] \d r = 0\ ,\nonumber \\
&&\left[ {d \ov du} \left({f r_{\rm cl}^2 \ov F^{1/2}} {d \ov du} \right)
+ \omega^2 {h r_{\rm cl}^2 F^{1/2} \ov g } \right] \d\th = 0\ ,\\
&&\left[ {d \ov du} \left({f r_{\rm cl}^2 \ov F^{1/2}} {d \ov du} \right)
+ \omega^2 {h r_{\rm cl}^2 F^{1/2} \ov g } \right] \d\phi = 0\ .\nonumber
\ea
For the straight strings the expressions for the action and the equations of motion were given
in \cite{SS} and will not be repeated here.
As discussed extensively in \cite{ASS1,SS} instabilities are indicated by the presence of a
normalizable zero-mode which can exist for specific values of $u_0$, denoted by $u_{0c}$.
Such values are in general different than possible values of $u_0$ at which the length
reaches an extremum, denoted by  $u_{0m}$ and found by solving  $L'(u_0)=0$.
Our aim is to find for the zero mode the critical curve in the parametric space of
$a$ and $u_0$ which separates the stable from the unstable regions.

\no
Finally, we turn to the fluctuations of the $D5$-brane and
perturb the embedding according to
\ba
\label{3-5}
x^\m=\d x^\m(t,\th_\a)\ , \qq u=u_0\ ,\qq x^\m = x,y,z\ ,
\ea
leaving the position of the $D5$-brane at $u=u_0$ intact due to the gauge choice $u=\s$ for the
strings.
The second order, in the fluctuations, term in the expansion
of the $D5$-brane action is
\ba
\label{3-6}
S={NR\ov 8\pi^4}\int dtd\Om_5\sqrt{\g}\sqrt{-G_{tt}}
\bigg\{1+
{G_{\m\n}\ov 2}\g^{\a\b}\del_\a\d x^\m\del_\b\d x^\n
+{G_{\m\n}\ov 2G_{tt}}\d\dot{x}^\m\d\dot{x}^\n\bigg\} \ ,
\ea
where $\g_{\a\b}$ is the metric of $S^5$ and the action is calculated at $u=u_0$. The
subscripts $\a,\m$ refer to the angles of $S^5$ and to the $x,y,z$ coordinates, respectively.
We expand the fluctuations in terms of
the spherical harmonics of the $S^5$-sphere $\Psi_\ell$, satisfying the eigenvalue eq.
$\nabla_{\g}^2\Psi_\ell=-\ell(\ell+4)\Psi_\ell$, $\ell=0,1,2,\dots$,
as
\be
\d x^\m(t,\th_\a) = \d x^\m(t) \Psi_\ell(\th_\a)\ .
\ee
Then, from the Euler--Lagrange equations for the action we find that
\ba
\label{3-8}
{d^2\d x^\m\ov dt^2}+\Om_\ell^2\d x^\m=0\ , \quad \Om_\ell^2=-G_{tt}(u_0)\ell(\ell+4)>0\ .
\ea
Thus the classical solution is stable under the $\d x^\m$ fluctuations. Note that, there are no
boundary conditions for these fluctuations, the reason being that the $\mathbb{R}\times S^5$
has no boundary.

\subsection{Boundary and matching conditions}

To fully specify our eigenvalue problem, we must impose boundary conditions on the string
fluctuations in the UV limit, $u\to\infty$ and in the IR limit, $u=u_{\rm min}$, as well as
matching conditions at the baryon vertex, i.e. $u=u_0$. The boundary condition at the UV are
chosen so that the quarks are kept fixed, that is
\ba
\Phi(u) = 0\ ,\qq {\rm as}\quad u\to \infty \ ,
\label{3-9}
\ea
where $\Phi$ is any of the fluctuations of any of the $k$-strings.
At $u=u_{\rm min}$ we demand finiteness of the solution
and its $u$-derivative for the $N-k$ straight strings that extend in there,
so that the perturbative approach is valid. It turns out that
for the zero-mode they completely decouple from the analysis, as
in the case of the quark-monopole potential in \cite{SS}.
Since we consider a uniform distribution of quarks on
a spherical shell, the $D5$-brane will be located by symmetry at $x=y=z=0$.
For each one of the $k$-strings,
$\d r$ is its longitudinal fluctuation and $\d\th,\d\phi,$ are its transverse ones.

\no
For the longitudinal fluctuations $\d r$, it
is more convenient to work out the necessary boundary condition
in the gauge $r=\s$, instead of $u=\s$ and then translate the result in the latter gauge
where the equations of motion are simpler.
In that spirit, following also an analogous treatment in  \cite{ASS1,SS},
the gauge change is effectively done by the coordinate transformation
\ba
\label{3-10}
u=\overline{u}+\d u(t,u)\ ,\qq \d u(t,u)=-{\d r(t,u)\ov r_{\rm cl}^{\prime}}\ .
\ea
\no
Hence, the boundary condition at the baryon vertex can be found, if we temporarily change
gauge to $r=\s$ and apply small fluctuations for the $\bar u$-coordinate. Demanding
a well behaved boundary problem for $\d u$-fluctuations we obtain
\ba
\label{3-11}
{d\d u\ov d r}\bigg{|}_{u=u_0}=0\ .
\ea
From \eqn{2-10}, \eqn{3-10} and \eqn{3-11} we find
that the $\d r$-fluctuations, for each of the $k$-strings, satisfy the following boundary condition
\ba
\label{3-12}
u=u_0:\qq 2(f-f_1)\d r^{\prime}
+\d r\left(2f^{\prime}-{f^{\prime}\ov f}f_1-{g^{\prime}\ov g}(f-f_1)\right)=0\ .
\ea
This is an equation for $u_0$ in terms of $a$ and the
parameters of the specific problem. We will denote its solution, whenever it exists, by $u_{0c}$.
We will not assume continuity relations between the fluctuations of the $D5$-brane
and the $N$-strings.

\no
In general, the value of $u_{0c}$ does not coincide with
that corresponding to a possible extremum of the
radius $L(u_0)$, denoted by $u_{0m}$. For the
quark-antiquark potential these values coincide \cite{ASS1,ASS2}, but
they do not do so for the case of the quark-monopole potential \cite{SS}.
In our case using \eqn{2-12}, \eqn{2-15} and \eqn{2-16} we find that
\ba
\label{3-16}
&&{dE\ov du_0}={k\sqrt{f_1}\ov2\pi}{dL\ov du_0}\ , \nonumber \\
&&{dL\ov du_0}= - {\sqrt{f_1F}\ov f}+{\del u_1\ov\del u_0}\ {f_1^{\prime}\ov2\sqrt{f_1}}
\int_{u_0}^\infty du{\sqrt{gf}\ov(f-f_1)^{3/2}}\ ,
\ea
where from \eqn{2-16} we compute that
\be
f_1' {\del u_1\ov\del u_0} = \sin^2\Th f_0'-{R\ov 2 a} f_0 \cos \Th
\del_u\left({\del_u \sqrt{-G_{tt}}\ov \sqrt{g}}\right)\! \bigg |_{u=u_0}\ .
\label{f1u1}
\ee
A necessary condition for $u_{0m}$ to exist is that the right hand side of
\eqn{f1u1} is positive,
which is indeed satisfied in
the non-extremal $D3$-branes example we will examine in the next section.
Also the extrema of $L(u_0)$ and $E(u_0)$ at $u_0=u_{0m}$
coincide as in the case of the quark-antiquark
and quark-monopole potential energy interactions
shown in \cite{ASS2} and \cite{SS}, respectively.
Similarly to \cite{bs}, we can easily show that
\ba
\label{3-14}
&&{dE\ov dL}={k\ov2\pi}\sqrt{f_1}\ ,\nonumber \\
&&{d^2E\ov dL^2}={k\ov4\pi}{f_1^{\prime}\ov\sqrt{f_1}}\ {1\ov L^{\prime}(u_0)}\
{\del u_1\ov\del u_0}\ .
\ea
Since $dE/dL>0$ the force is manifestly attractive. On the other hand its magnitude
may decrease or increase with distance depending on the sign of the right had side of the
expression for $d^2E/dL^2$. Note that this is not necessarily in conflict with gauge theory
expectations since there is no analog for the binding energy of baryons
of the concavity condition of the heavy quark-antiquark pair
found in \cite{concavity}.
In addition, having assumed that the quarks, the baryon consists of, are located on a
spherical shell has an inherit degree of arbitrariness that is reflected in the
details of the expressions we have derived.

\subsection{Zero modes}

We will now study the zero mode problem for
a spherical shell distribution and find the critical curve in the
parametric region of $(a,u_{0})$.

\no
The longitudinal and transverse fluctuations of each string,
satisfying already the conditions in the UV, are
\ba
\label{3-13}
&&\d r=A\ J(u)\ ,\qq \d\th=B\ K(u)\ ,\qq \d\phi=C\ K(u)\ ,
\nonumber\\
&&J(u)=\int^{\infty}_u du {\sqrt{g f}\ov (f-f_1)^{3/2}}\ ,
\qq K(u)=\int^{\infty}_u {du\ov f r_{\rm cl}^2}\ \sqrt{gf\ov f-f_1}\ ,
\ea
where $A,B$ and $C$ are integration constants.
Using \eqn{2-3} and \eqn{2-10} we see that $K(u)$ diverges, thus we should choose $B=C=0$.
Therefore, the
transverse fluctuations are stable as in the cases of the quark-antiquark and
quark-monopole potentials \cite{ASS1,SS}.

\no
For the longitudinal fluctuations we should substitute the above expression for $\d r$
into \eqn{3-12} and obtain a transcendental equation that defines the critical curve
separating the stable from the unstable regions.

\section{Examples}

In this section we will first review the behavior of the baryon potential emerging in the
conformal case \cite{Brandhuber:1998,Imamura1} and examine
its stability behavior. Then we will present the analogous analysis
for the finite temperature case.

\subsection{The conformal case}
\subsubsection{Classical Solution}
At first we will consider the conformal case, in which
the metric of $AdS_5\times S^5$ has the form of \eqn{2-1}, with
\be
\label{4-1}
-G_{tt}=G_{xx}= G_{uu}^{-1}={u^2\ov R^2}\ .
\ee
The radius and the energy corresponding to a uniform distribution of quarks on a spherical shell
are given in terms of the
position of the $D5$-brane $u_0$ and the turning point $u_1$ of each string
\ba
\label{4-2}
L={R^2u_1^2\ov 3u_0^3}\ {\cal I}\ ,\qq E={k u_0\ov 2\pi}\left(- {\cal J}+{5-4a\ov 4a }\right)\ ,
\ea
with
\be
{\cal I}= {_2F_1}\left({1\ov 2},{3\ov 4},{7\ov 4};{u_1^4\ov u_0^4}\right)\ ,\qq
{\cal J}={_2F_1}\left(-{1\ov 4},{1\ov 2},{3\ov 4};{u_1^4\ov u_0^4}\right)\
\label{uinnt1}
\ee
and where ${_2F_1}(a,b,c;z)$ is the hypergeometric function and $u_0\geqslant u_1\geqslant 0$.
From \eqn{2-16} we find that in this case the no-force condition on the $u$-axis yields
\ba
\label{4-3}
u_1=u_0(1-\lambda^2)^{1/4}\ ,\qq \l={5-4a\ov 4a}\ .
\ea
We note that \eqn{4-3} does not restrict the domain of $u_0$, but this will not be the case
when temperature is turned on.
Since $\l <1 $, a baryon configuration exists if
$a> a_<$, $a_<=5/8=0.625$. This value was found in \cite{Brandhuber:1998,Imamura1}.
Using \eqn{4-2} and \eqn{4-3} we can find the binding energy in terms
of the physical length
\ba
\label{4-4}
E=-{R^2\ov2\pi L}{k\sqrt{1-\l^2}\ov 3}\left({\cal J}-{5-4a\ov 4 a}\right){\cal I}\ .
\ea
From this expression it is apparent why the possibility $a=5/8$ ($\l=1$) is discarded.
The energy of the baryon goes as $1/L$ as it was expected by conformal invariance and
has the familiar for these type of computations dependence on the 't Hooft coupling.

\subsubsection{Stability analysis}

Next we consider the stability behavior of the conformal case.
For the longitudinal fluctuations we have to solve \eqn{3-12},
where
\ba
\label{4-5}
\d r(u)= A \int_u^\infty du{u^2\ov(u^4-u_1^4)^{3/2}}={A\ov 3u^3}\ {_2F_1}
\left({3\ov 4},{3\ov 2},{7\ov 4};{u_1^4\ov u^4}\right)\ .
\ea
Substituting \eqn{4-3} and \eqn{4-5} in \eqn{3-12}
we find the following transcendental equation for the zero mode to exist
\ba
\label{4-6}
{_2F_1}\left({3\ov4},{3\ov2},{7\ov4};1-\l^2\right)={3\ov2\l(1+\l^2)}\ .
\ea
Using \eqn{4-3} and \eqn{4-6} we find numerically a critical value for $a$,
i.e. $a_c\simeq 0.813$\ .

\no
Thus, the stability analysis sets a low bound for $a$ which is still less than unity.
However,
given that this corresponds to the conformal point of a maximally
$\cN=4$ supersymmetric theory we find it remarkable. The only other case
of perturbative instability in the conformal
point is for the quark-antiquark potential in the $\beta$-deformed $\cN=1$ theory \cite{LS},
whose
supergravity dual was constructed in \cite{LM}. In this case
the instability occurs for values of the $\s$-deformation parameter larger than
a critical value \cite{ASS2}.

\subsection{Non--extremal D3--branes}
\subsubsection{Classical solution}

In this section we consider a stack of $N$ non-extremal $D3$--branes \cite{HoroStro}.
In the field theory limit
the metric has the form of \eqn{2-1} where
\ba
\label{4-7}
G_{tt}= -G_{uu}^{-1}=-{u^2\ov R^2}\left(1-{\mu^4\ov u^4}\right)\ ,\qq G_{xx}={u^2\ov R^2}\ .
\ea
The corresponding Hawking temperature is $T_H=\m/(\pi R^2)$.
In what follows, we will compute in units of $\m$, so that we set $\m=1$.
The radius and the energy read
\ba
\label{4-8}
&&L=R^2{\sqrt{u_1^4-1}\ov 3u_0^3}
F_1\left({3\ov 4},{1\ov 2},{1\ov 2},{7\ov 4};{1\ov u_0^4},{u_1^4\ov u_0^4}\right)\ ,
\nonumber \\
&&E={k\ov 2\pi}\left\{{\cal E}+{u_0\ov 4a}\sqrt{1-{1\ov u_0^4}}+{1-a\ov a}(u_0-1)\right\}\ ,
\\
&&{\cal E}=-u_0 F_1\left(-{1\ov 4},-{1\ov 2},{1\ov 2},{3\ov 4};{1\ov u_0^4},{u_1^4\ov u_0^4}\right)
+1\ ,
\nonumber
\ea
where $F_1(a,b_1,b_2,c;z_1,z_2)$
is the Appell hypergeometric function and $u_0\geqslant u_1\geqslant 1$.
From \eqn{2-16} we find that in this case the no-force condition on the $u$-axis yields
\ba
\label{4-9}
\cos\Th={1-a\ov a}+{1+1/u_0^4\ov 4a\sqrt{1-1/u_0^4}}\ ,
\qq \cos\Th=\sqrt{{u_0^4-u_1^4\ov u_0^4-1}}\ .
\ea
Employing the fact that $\cos\Th\leqslant1$ in \eqn{4-9}
we end up with a second order polynomial inequality which
has a solution for $a\geqslant5/8$, namely
\ba
\label{4-10}
&&u_0\geqslant u_<(a)\ ,\qq u_<(a)=1/\l^{1/4}\ ,
\nonumber \\
&&\l=-9-32a(a-1)+4\sqrt{2}(2a-1)\sqrt{3-8 a(1-a)}\ ,
\\
&&a\in[5/8,1]\ ,\quad \l\in[0,-9+4\sqrt{6}]\ ,\quad u_0\in[\infty, u_<(1)]\ ,
\quad u_<(1)\simeq 1.058\ .
\nonumber
\ea
However, if $a=5/8$ there is no classical solution since $u_0\to\infty$.
For $a>5/8$ the solution of \eqn{4-9} is
\ba
\label{4-11}
u_1=u_0(1-\L^2(u_0))^{1/4}\ ,\qq \L(u_0)={1-a\ov a}\sqrt{1-1/u_0^4}+{1+1/u_0^4\ov 4a}\ .
\ea
\no
In this case we can not express $E$ in terms of $L$.
In fact $E$ is a doubled-valued function of $L$
with energetically favorable and unfavorable branches, as depicted in Fig. 2a.
It is similar to the quark-antiquark potential for the finite temperature case
\cite{wilsonloopTemp}.

\begin{figure}[!t]
\begin{center}
\begin{tabular}{cc}
\includegraphics[height=5.0cm]{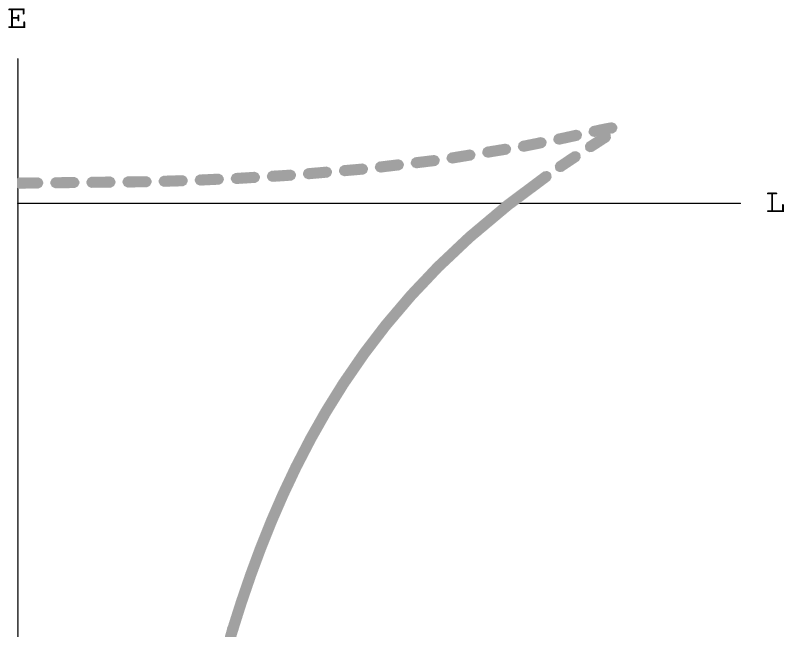}
&\includegraphics[height=6.6cm]{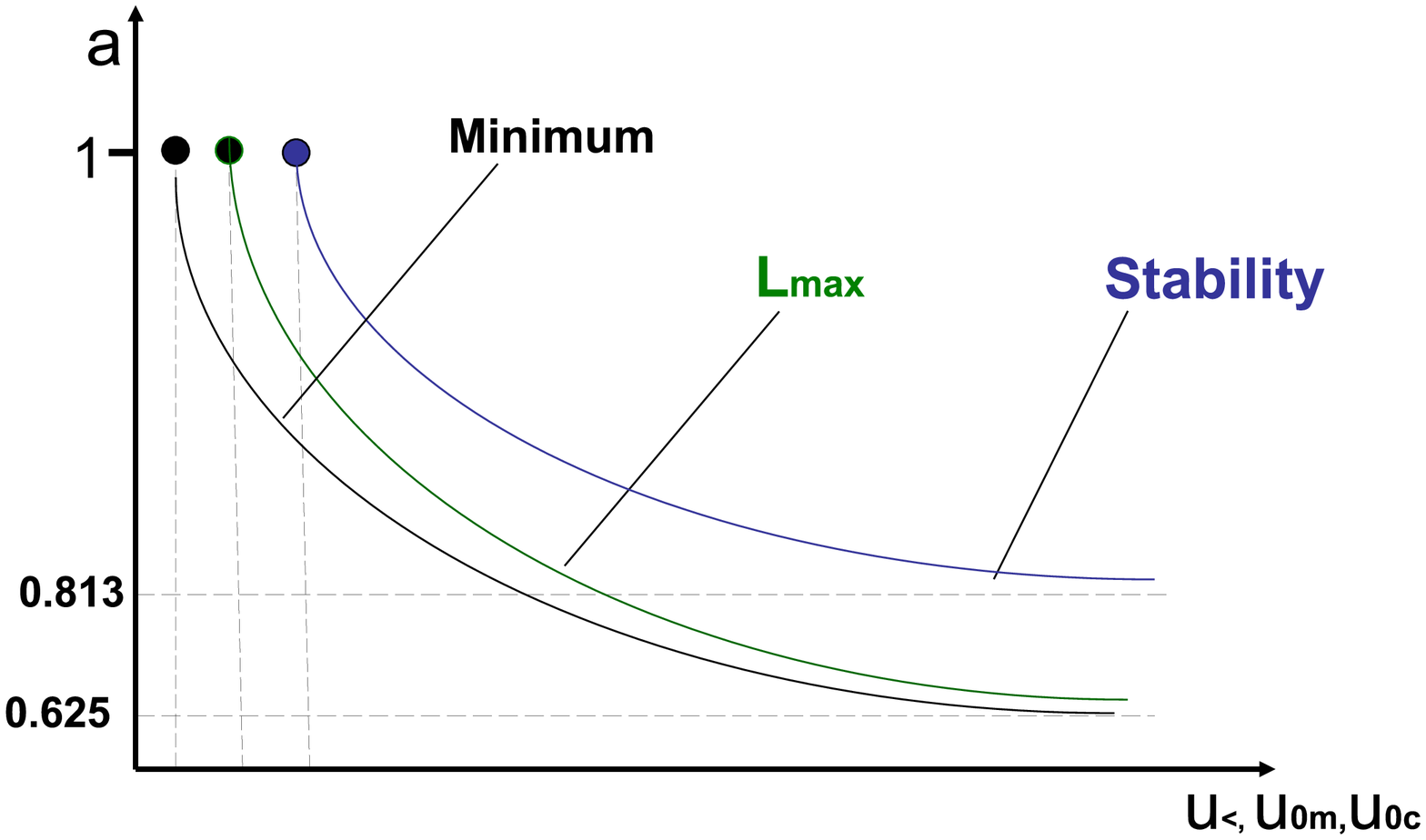}\\
(a) & (b)
\end{tabular}
\end{center}
\vskip -.4 cm \caption{(a) Plot of $E(L)$ for non extremal $D3$--branes. The various types of line correspond to stable
(solid grey) and unstable (dashed grey) configurations. As $a$ decreases unstable part extends to the lower branch as well.
This behavior is described in section (4.2.2).
(b) From the left to the right: i) Plot of $u_<$  versus a, ii) Plot $a$ versus $u_{0m}$,
iii) Plot of $u_{0c}$ versus a. The regions on the right and left hand side of the stability line
are stable and unstable, respectively.}
\end{figure}

\subsubsection{Stability analysis}

Next we will consider the stability behavior of the non-extremal
$D3$-branes. For the longitudinal fluctuations we have to solve
\eqn{3-12} where
\ba
\label{4-12} \d r(u)=A\int_u^\infty du
{\sqrt{u^4-1}\ov(u^4-u_1^4)^{3/2}}= {A\ov 3u^3}F_1\left({3\ov
4},-{1\ov 2},{3\ov 2},{7\ov 4};{1\ov u^4},{u_1^4\ov u^4}\right)\ .
\ea
From \eqn{3-12} and \eqn{4-12} we find that the zero
mode condition is given by the transcendental equation
\ba
\label{4-13} {3\ov
2\sqrt{u_0^4-u_1^4}}={2u_0^4-u_1^4-1\ov(u_0^4-1)^{3/2}}
F_1\left({3\ov 4},-{1\ov 2},{3\ov 2},{7\ov 4};{1\ov u_0^4},{u_1^4\ov u_0^4}\right)\ .
\ea
From numerical considerations
we find that for every value of $a$ larger than the critical value
found in the conformal case $a_c\simeq 0.813$ there is a zero mode
on the energetically favored branch of $E(L)$  i.e. $u_{0c}>u_{0m}$.
Its solution has the expansion
\be
\label{4-14} a_c\simeq 0.813+{0.679\ov u_{0c}^4}+{0.778\ov
u_{0c}^8}+{\cal{O}}\left(1\ov  u_{0c}^{12}\right)\ .
\ee
As $a_c$ decreases $u_{0c}$ increases and for $a_c\simeq 0.813$
even the conformal part becomes unstable.
The results of our stability analysis are summarized in Fig.2b.

\newpage
\no
Also, for the particular case of $a=1$ we present the table:
\begin{center}
\begin{tabular}{| c || c | c | c |}
\hline
{\bf Values for $a=1$} &{\bf Minimum} & {\bf Critical} & {\bf Maximum} \\
\hline\hline  $u_0$ & 1.0581 & 1.4292 & 1.3437 \\
\hline   $L/R^2$ & 0 & 0.2920 & 0.2951 \\
\hline  $E/N$ & 0.0097 & 0.0266 & 0.0273 \\
\hline
\end{tabular}
\end{center}
\no

\section{Magnetic flux and stability of the baryon vertex}

In this section we will generalize our discussion to the case where we turn on,
besides the electric ones,
non-trivial magnetic flux components in $F_{ij}$,
as in \cite{Lozano:2006}.
The fact that this is possible could be seen, from the fact that the $S^5$-sphere can
be regarded as
a $U(1)$ fibre over $\mathbb{CP}^2$ with a non-trivial fibre connection \cite{Pope:1980ub}.
This $U(1)$ connection, introduces a generalization of the baryon vertex in which non-zero magnetic
components of the Born--Infeld field $F_{ij}$ appear in the $\mathbb{CP}^2$ directions
in such a way that it
is self-dual with respect to that metric. For topological reasons, its flux over non-trivial
two-cycles of $\mathbb{CP}^2$ is quantized and $n^2$ represents the instanton number.
Working along similar lines as in section (2.2) of \cite{Lozano:2006}, we find that
the action of the $D5$-brane \eqn{2-5} in our conventions gets modified to
\ba
\label{5-1}
S_{D5}={TNR\sqrt{-G_{tt}}\ov 8\pi}\ \n \ ,
\ea
where
\be
\n \equiv  1+32\pi^2 b\ ,\qq b\equiv{n^2\ov R^4}\ .
\ee
Note that the relevant parameter is not the instanton number $n^2$ but its ratio with the
\\'t Hooft coupling.
The analogous expression to \eqn{2-15} for the binding energy of the baryon is
\ba
\label{5-2}
E = {k \ov 2\pi}\left\{\int_{u_0}^\infty du\sqrt{F}
- \int_{u_{\rm min}}^\infty du\sqrt{g}+
{1-a\ov a} \int_{u_{\rm min}}^{u_0}du \sqrt{g} + \n\
{R\sqrt{-G_{tt}}\ov 4 a }\bigg|_{u=u_0}\right\}\ ,
\ea
while the expression for the radius of the baryon is given by the same equation as in \eqn{2-12}.
The analogous expression to \eqn{2-16} for the no-force condition on the $u$-axis yields\footnote{
The conditions \eqn{3-16} and \eqn{3-14}
 are also valid in this case as well. In using \eqn{f1u1}
one should include an overall factor
of $\n$ to the second term in the right hand side.}
\ba
\label{5-3}
&&\cos\Th = {1-a\ov a} +
\n\ R {\del_u\sqrt{-G_{tt}}\ov 4a\sqrt{g}}\bigg{|}_{u=u_0}\ ,\nonumber\\
&&\cos\Th\equiv \sqrt{1-f_1/f_0}\ .
\ea
The stability analysis of this classical solution is easy to discuss.
Since the Born--Infeld field is self-dual,
the determinant under the square root in the action of the $D5$-brane
\eqn{2-5} is a perfect square, and
there no mixing terms of the induced metric of the $D5$-brane and the Born--Infeld field.
Furthermore, the Born--Infeld field is a two-form on $\mathbb{CP}^2$ that remains unchanged since
perturbing
the position of the $D5$-brane we keep the gauge choice \eqn{2-13}.
Hence, the stability analysis described in section 3, is applicable for non-zero
magnetic flux as well.

\no
We will next specialize to the conformal and non-extremal D3-branes cases.

\subsection{The conformal case}

We will first consider the classical solution for the conformal
case. The whole treatment can be done in terms of an effective
parameter $a_{\rm eff}$ (defined below) and using our previous
results in the absence of magnetic charge. The analogous expression
to \eqn{4-3} reads
\ba
\label{5-4} u_1=u_0(1-\l_{\rm eff}^2)^{1/4}\ ,\qq \l_{\rm
eff}={5-4a_{\rm eff}\ov 4a_{\rm eff}}\ , \qq a_{\rm eff}\equiv{a\ov
1+{32\pi^2 b\ov 5}}\ .
\ea
Since $\l_{\rm eff}< 1 $ a baryon
configuration exists for
\ba
\label{5-5} a_{\rm eff} > 5/8\quad
\Longrightarrow \quad  a >  a_<\ ,
\qq a_<={5\ov8}+4\pi^2 b \ .
\ea
However, since $a \leqslant 1$ we get a bound for the instanton
number, namely that
\ba
\label{5-6} a_< < 1\quad
\Longrightarrow\quad {n^2\ov R^4} < {3\ov32\pi^2}\simeq 0.0095\ ,
\ea
which is the bound found in
\cite{Lozano:2006} for a classical configuration to exist.
The
binding energy in terms of the radius reads
\ba
\label{5-7}
E=-{R^2\ov2\pi L}{k\sqrt{1-\l_{\rm eff}^2}\ov 3}\left({\cal J}-
{5-4a_{\rm eff}\ov 4 a_{\rm eff}}\right){\cal I}\ ,
\ea
where ${\cal I}$ and ${\cal J}$ were given in \eqn{uinnt1}.
We have shown in section 4 that
there is a zero mode for $a_{\rm eff}\simeq 0.813$, for which the
system becomes unstable. This improves the above bound for the
instanton number, in comparison to the 't Hooft coupling, to
\be
{n^2\ov R^4 } \lesssim 0.0036\ ,
\ee that
should be respected for the classical configuration not only to
exist, but also to be perturbatively stable. Stability analysis sets
lower and higher bounds for the instanton number and $a$, $0.0036$
and $0.813$ respectively. This behavior is depicted in Fig.3.

\begin{figure}[!t]
\begin{center}
\begin{tabular}{cc}
\includegraphics[height=10.0cm]{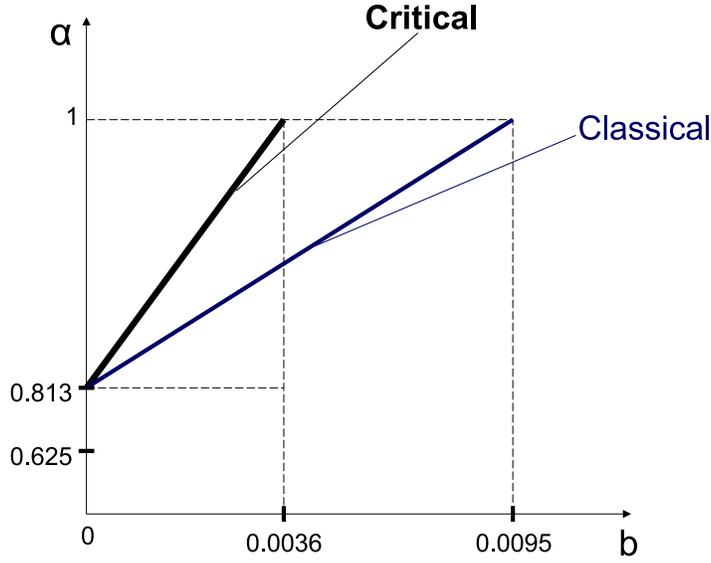}
\end{tabular}
\end{center}
\vskip -1.5 cm \caption{Classical and critical lines of $a$ versus $b$, i.e.
$a_{\rm eff}=5/8,\ 0.813$, respectively.
There is a classical solution for values of $(a,b)$, on the right hand side
of the classical line. As for stability,
the regions on the left and the right hand sides of the critical line
are stable and unstable, respectively. }
\end{figure}

\subsection{Non-extremal $D3$-branes}

Finally, we will consider the non-extremal $D3$-branes. The
analogous expression to \eqn{4-8} for the energy read
 \ba
\label{5-9}
&&E={k\ov 2\pi}\left\{{\cal E}+{1-a\ov a}(u_0-1)+ \n {u_0\ov 4a}\sqrt{1-{1\ov u_0^4}}\
 \right\}\ ,
\nonumber
\\
&&{\cal E}=-u_0 F_1\left(-{1\ov 4},-{1\ov 2},{1\ov 2},{3\ov 4};{1\ov u_0^4},{u_1^4\ov u_0^4}\right)
+1\ .
\ea
The analogous
expression to \eqn{4-9} no-force condition, has a solution for
$\displaystyle{a > {4+\n\ov8}}$, namely that
\ba
\label{5-10}
&&u_1=u_0(1-\L^2(u_0))^{1/4}\ ,\qq \L(u_0)={1-a\ov a}\sqrt{1-1/u_0^4}+ \n {1+1/u_0^4\ov 4a}\ ,
\nonumber\\
&&u_0\geqslant u_<\ , \qq u_<=1/\l^{1/4}\ ,
\\
&&\l={1\ov\n^2}\left(-8-\n^2+32(-a+1)a+4\sqrt{2}(2a-1) \sqrt{2-8a(1-a)+ \n^2}\ \right)\ .\nonumber
\ea
The energy versus radius is a doubled valued function
with an energetically favorable and an energetically unfavorable
branch, as depicted in Fig.2a.

\no
Turning now to the stability behavior of the classical solution,
we found that for every value of $a$ and instanton number above
the critical line of Fig.3, there is always a zero-mode on the
energetically favorable branch of $E(L)$, giving rise to a
critical curve bounded from below by the conformal
critical line.
As $a$ and/or the instanton number reach
the critical line, even the conformal part of the diagram becomes unstable.


\section{Summary and Discussion}

In this paper we first constructed classical configurations with strings and a $D5$-brane that
represent bound, baryon-like states, within the gauge/gravity correspondence
and for a general class of
backgrounds. We worked out explicit examples in the $\cN=4$ SYM for $SU(N)$
at the conformal point and at finite temperature.
Stability restricts the value of $k$ to be larger than a critical value
even in the conformal case ($0.813 N$, higher than the classical lower bound $0.625 N$).
It addition, a part of the energetically favorable branch of
the multi-branch potential arising in the finite temperature case should be disregarded as unstable,
even if $k=N$.
If the baryon vertex is supplemented by magnetic charge, we showed
that stability requires un upper bound for the instanton number, as compared to the 't Hooft coupling,
associated with a self-dual gauge field.
This is unlike the findings for quark-antiquark potentials in \cite{ASS1,ASS2}
where stability discards energetically unfavorable branches completely
and for the dyon-dyon potentials in \cite{SS}, where parts
of the energetically unfavorable branch become perturbatively stable.

\no
One may well wonder how generic is our discussion in the finite temperature case, having
used a specific black $D3$-brane background. We have checked
that for the Rindler space, that represents the horizon of any black hole-type solution
similar phenomena occur.
There is a critical value of $k$ below which
there is instability and also part of the energetically favorable branch is unstable.
Hence, our results are quite generic for a large class of backgrounds with black hole-type behavior.
Moreover, our
techniques could be extendable to theories with dynamical flavor,
such as the one in \cite{Casero:2006pt}.

\no
Our work is also straightforwardly extendable to supergravity backgrounds
in which off-diagonal metric elements appear.
This cover cases with non-vanishing Wess--Zumino term and cases of moving baryons
in hot quark-gluon plasmas.
The classical solution of a moving baryon in hot plasma was studied in \cite{Liu:2008}.
The end result is that for a uniform circle distribution at the boundary of $AdS_5$,
the binding energy versus length has two branches and for high enough velocities
$L_{max}\sim(1-v^2)^{1/4}/T_H$,
where the proportionality coefficient depends on the direction that the baryon moves,
with respect to the axis of the circle.
We have checked the stability behavior and
found that for perpendicular to the
circle distribution motion, there is an instability on the energetically favored
branch and as speed increases and/or $a$ decreases,
even the conformal part of the diagram becomes unstable.
We expect this behavior
to be valid for arbitrary quark distributions in baryons and moving directions.
It is quite interesting if this is further pursued.

\no
One could wonder if there are cases when any number of quarks below $k=N$ gives rise to
unstable configurations, thus realizing our original motivation that
prompted this work, i.e. only colorless
states can exist. It is not a surprise that this does not seem to happen for the $\cN=4$ SYM theory,
since
after all this is a maximally supersymmetric finite theory.
It will be interesting to explore this issue
in other theories with reduced  supersymmetry.

\no
In our investigation we have chosen to work for simplicity
with a quark distribution in the baryon that is homogeneous and confined on a spherical shell.
This choice is quite arbitrary and although it preserves radial symmetry it is not
a priory the one that the quarks would prefer if, for instance, we apply a variational
principle that minimizes the total energy of the configuration. It would be very interesting
to find the mechanism that determines from first principles the appropriate quark distribution
in a baryon within this framework. As a last comment on that, we have checked
that a randomly chosen radially symmetric distribution of quarks in the baryon
is inappropriate since
in that case, not even the classical configuration exists, i.e. the
no-force condition at the baryon vertex is not satisfied.

\centerline{\bf Acknowledgments}

\no
We acknowledge support provided through the European Community's program
''Constituents, Fundamental Forces and
Symmetries of the Universe" with contract MRTN-CT-2004-005104 and a program with
contract ANR-05-BLAN-0079-02.\\
In addition, K. Sfetsos acknowledges support provided by the
Groupemeur d' Int\'er\^et scientifique
P2I (physique des deux infinis).\\
In addition, K. Siampos acknowledges support provided by the
Greek State Scholarship Foundation (IKY). Also
would like to thank the
department of Mathematics of Kings
College London for hospitality and support provided by the "European Superstring
Theory Network" with contract MCFH-2004-512194, where part of this work
was done.

\end{document}